\documentclass[lettersize,journal]{IEEEtran}

\usepackage[colorlinks,urlcolor=black,linkcolor=blue,citecolor=blue]{hyperref}

\def\BibTeX{{\rm B\kern-.05em{\sc i\kern-.025em b}\kern-.08em
    T\kern-.1667em\lower.7ex\hbox{E}\kern-.125emX}}

\usepackage{csquotes}
\usepackage{rotating}
\usepackage{amssymb}
\usepackage{algorithm}
\usepackage{algpseudocode}
\usepackage{color}
\usepackage{epsf}
\usepackage{psfrag}
\usepackage{epsfig}
\usepackage{graphicx}
\usepackage{amsfonts}
\usepackage{textcomp}
\usepackage{comment}
\usepackage{footnote}
\usepackage{tablefootnote}
\usepackage[flushleft]{threeparttable}
\usepackage{paralist}
\usepackage{mdwlist}
\usepackage{amssymb}
\usepackage{amsmath}
\usepackage{url}
\usepackage{verbatim}
\usepackage{alltt}
\usepackage{balance}
\usepackage{multirow}
\usepackage{epstopdf}
\usepackage{lipsum}
\usepackage{float}
\usepackage{booktabs}
\usepackage{slashbox}
\usepackage[table]{xcolor}
\usepackage{array}
\usepackage{boldline}
\usepackage{caption,setspace}
\usepackage{mathtools}
\usepackage{csquotes}

\usepackage{stackengine}[2013-10-15]

\usepackage[T1]{fontenc}
\usepackage{frcursive}
\usepackage{calligra}
\usepackage{wedn}
\usepackage{bm}
\usepackage{aurical}
\usepackage{upgreek}

\definecolor{darkgreen}{RGB}{52,120,68}
\definecolor{blue1}{RGB}{0,15,133}

\usepackage{cite}
\usepackage{graphicx}
\usepackage{adjustbox}

\usepackage{esint}
\usepackage{tipa}

\newcolumntype{?}{!{\vrule width 1pt}}
\newcolumntype{+}{!{\vrule width 1.25pt}}

\def\hlineb#1{%
\noalign{\ifnum0=`}\fi\hrule \@height #1 %
\futurelet\reserved@a\@xhline}

\usepackage{cancel}

\ifCLASSINFOpdf
\else
\fi
\begin{document}

\title{Unraveling Movie Genres through Cross-Attention Fusion of Bi-Modal Synergy of Poster}

\author{Utsav Kumar Nareti,~\IEEEmembership{Student Member,~IEEE}, 
    Chandranath Adak,~\IEEEmembership{Senior Member,~IEEE},
    Soumi Chattopadhyay,~\IEEEmembership{Senior Member,~IEEE},
    Pichao Wang
\thanks{
U. K. Nareti and C. Adak are with 
the Dept. of CSE, Indian Institute of Technology Patna, Bihar 801106, India. 
S. Chattopadhyay is with 
the Dept. of CSE, Indian Institute of Technology Indore, Madhya Pradesh 453552, India.
P. Wang is with 
Amazon Prime Video, Seattle, Washington, United States-98121.
The work does not relate to P. Wang’s position at Amazon. 
Corresponding author: C. Adak (email: chandranath@iitp.ac.in)

This work has been submitted to the IEEE for possible publication. Copyright may be transferred without notice, after which this version may no longer be accessible.
}
}

\markboth{U. K. Nareti \MakeLowercase{\textit{et al.}}: Unraveling Movie Genres through Cross-Attention Fusion of Bi-Modal Synergy of Poster}
{U. K. Nareti \MakeLowercase{\textit{et al.}}: Unraveling Movie Genres through Cross-Attention Fusion of Bi-Modal Synergy of Poster}


\maketitle

\begin{abstract}
Movie posters are not just decorative; they are meticulously designed to capture the essence of a movie, such as its genre, storyline, and tone/vibe. For decades, movie posters have graced cinema walls, billboards, and now our digital screens as a form of digital posters. Movie genre classification plays a pivotal role in film marketing, audience engagement, and recommendation systems. Previous explorations into movie genre classification have been mostly examined in plot summaries, subtitles, trailers and movie scenes. Movie posters provide a pre-release tantalizing glimpse into a film’s key aspects, which can ignite public interest. In this paper, we presented the framework that exploits movie posters from a visual and textual perspective to address the multilabel movie genre classification problem. Firstly, we extracted text from movie posters using an OCR and retrieved the relevant embedding. Next, we introduce a cross-attention-based fusion module to allocate attention weights to visual and textual embedding. In validating our framework, we utilized 13882 posters sourced from the Internet Movie Database (IMDb). The outcomes of the experiments indicate that our model exhibited promising performance and outperformed even some prominent contemporary architectures.
\end{abstract}



\section{Introduction}\label{sec:intro}

\noindent
\IEEEPARstart{T}{he} film industry has experienced a significant transition from conventional theaters to online streaming/ OTT platforms like Amazon Prime Video, Netflix, YouTube, and Disney$^+$ due to the exponential growth of the Internet, and lockdowns during COVID-19 boosted it further.
Movie posters have evolved to this change. It was once designed primarily for large-scale displays in theaters or physical media. It must now effectively communicate its message in the digital posters or compact thumbnail format commonly used in OTT platforms. This transition requires a re-imagination of poster design, focusing on bold imagery, clear typography, and concise storytelling that can captivate viewers even on the limited screen of smartphones.

\begin{figure}[!t]
\centering
\scriptsize
\begin{adjustbox}{width=0.48\textwidth}
\begin{tabular}{c|c|c|c|c}
\includegraphics[width=0.18\linewidth, height=0.27\linewidth]{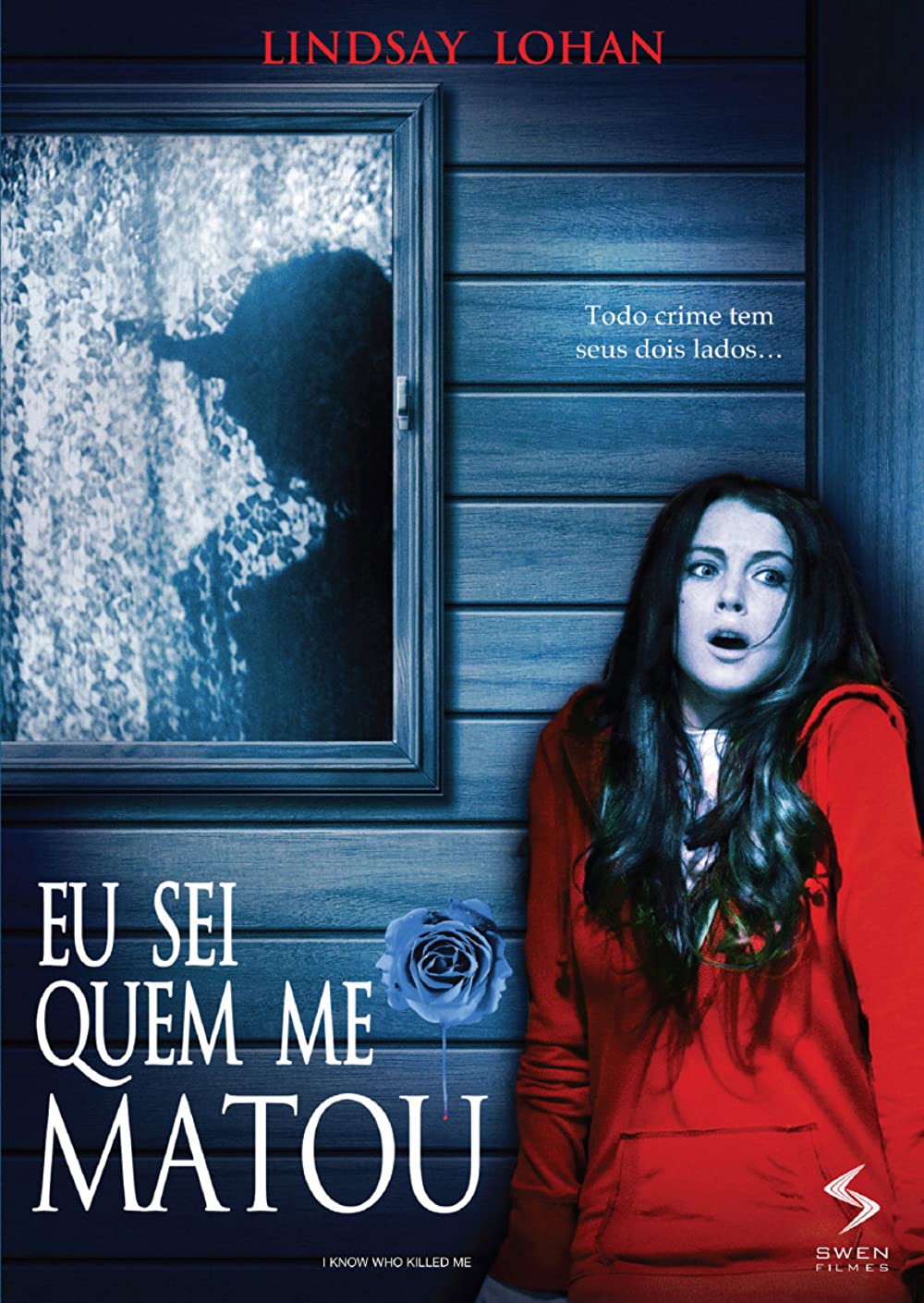} & 
\includegraphics[width=0.18\linewidth, height=0.27\linewidth]{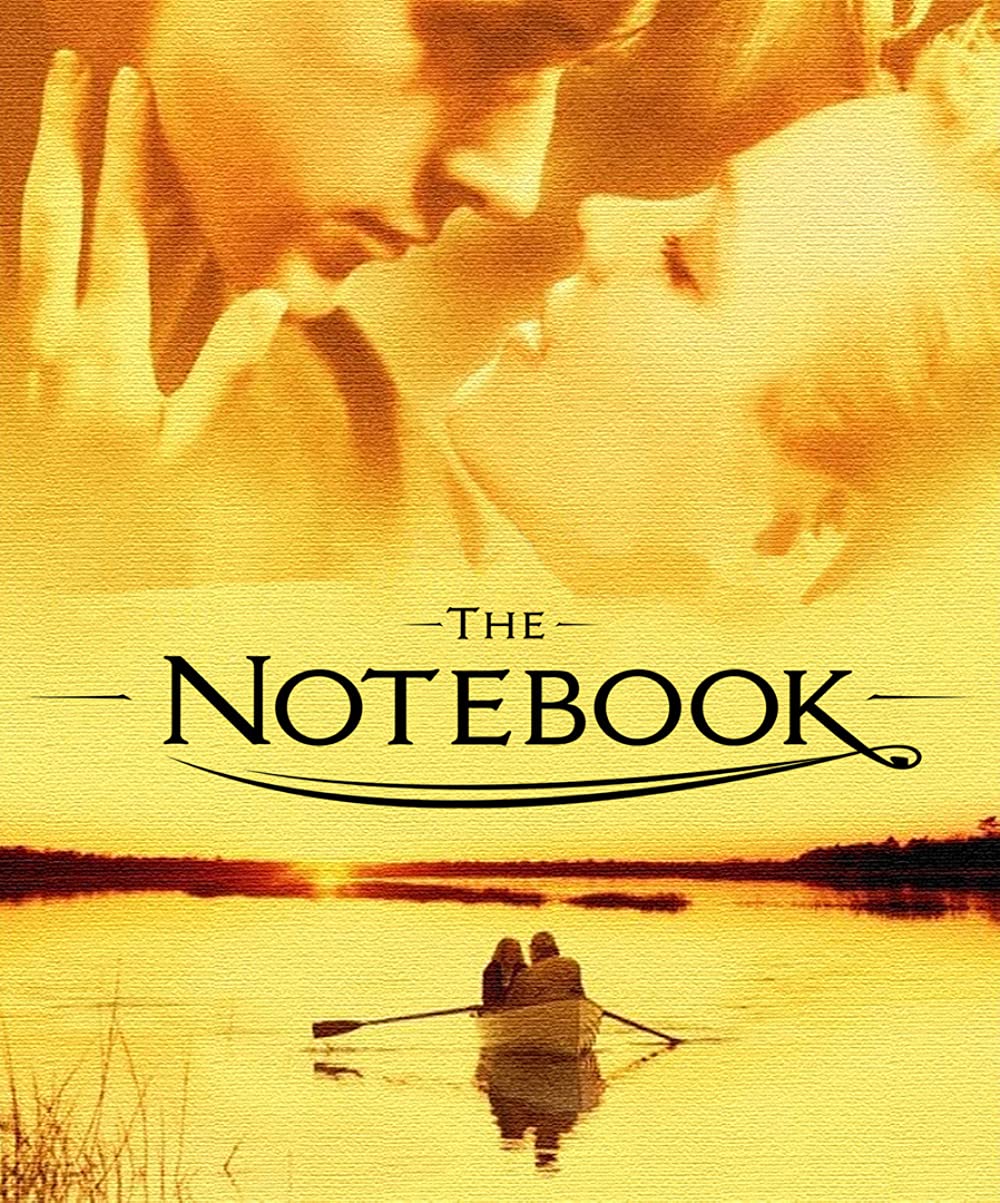} & 
\includegraphics[width=0.18\linewidth, height=0.27\linewidth]{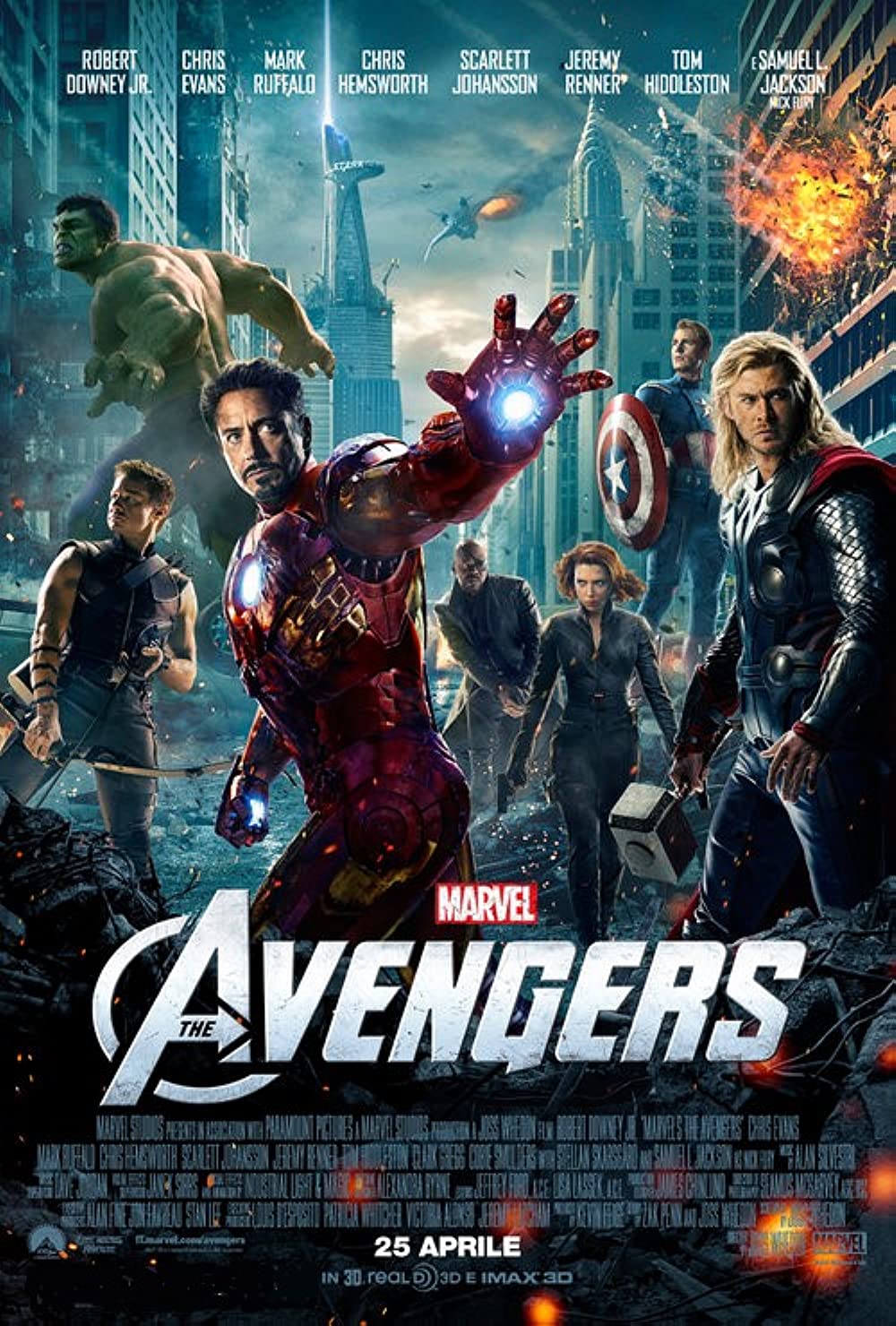} & 

\includegraphics[width=0.18\linewidth, height=0.27\linewidth]{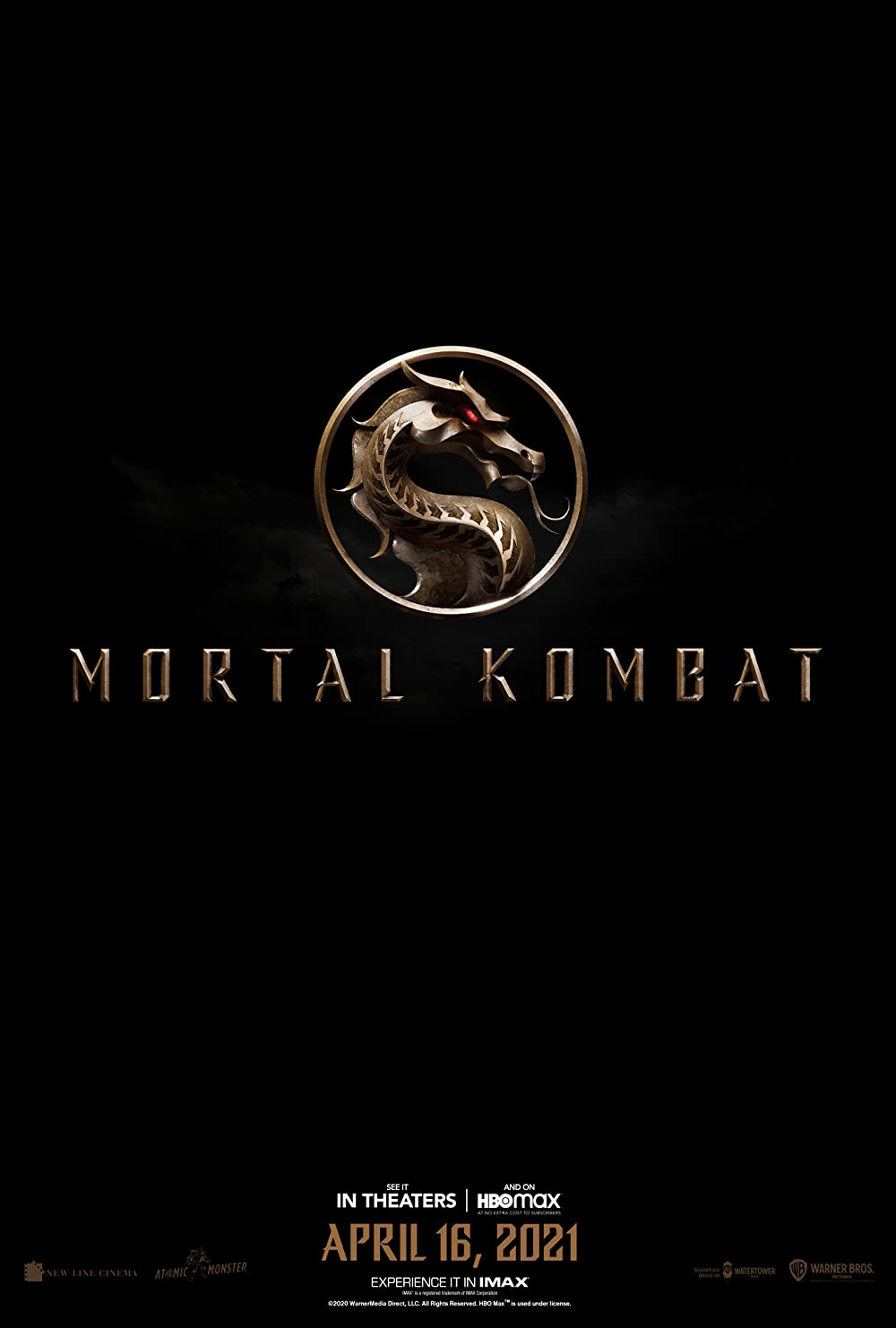} & 

\includegraphics[width=0.18\linewidth, height=0.27\linewidth]{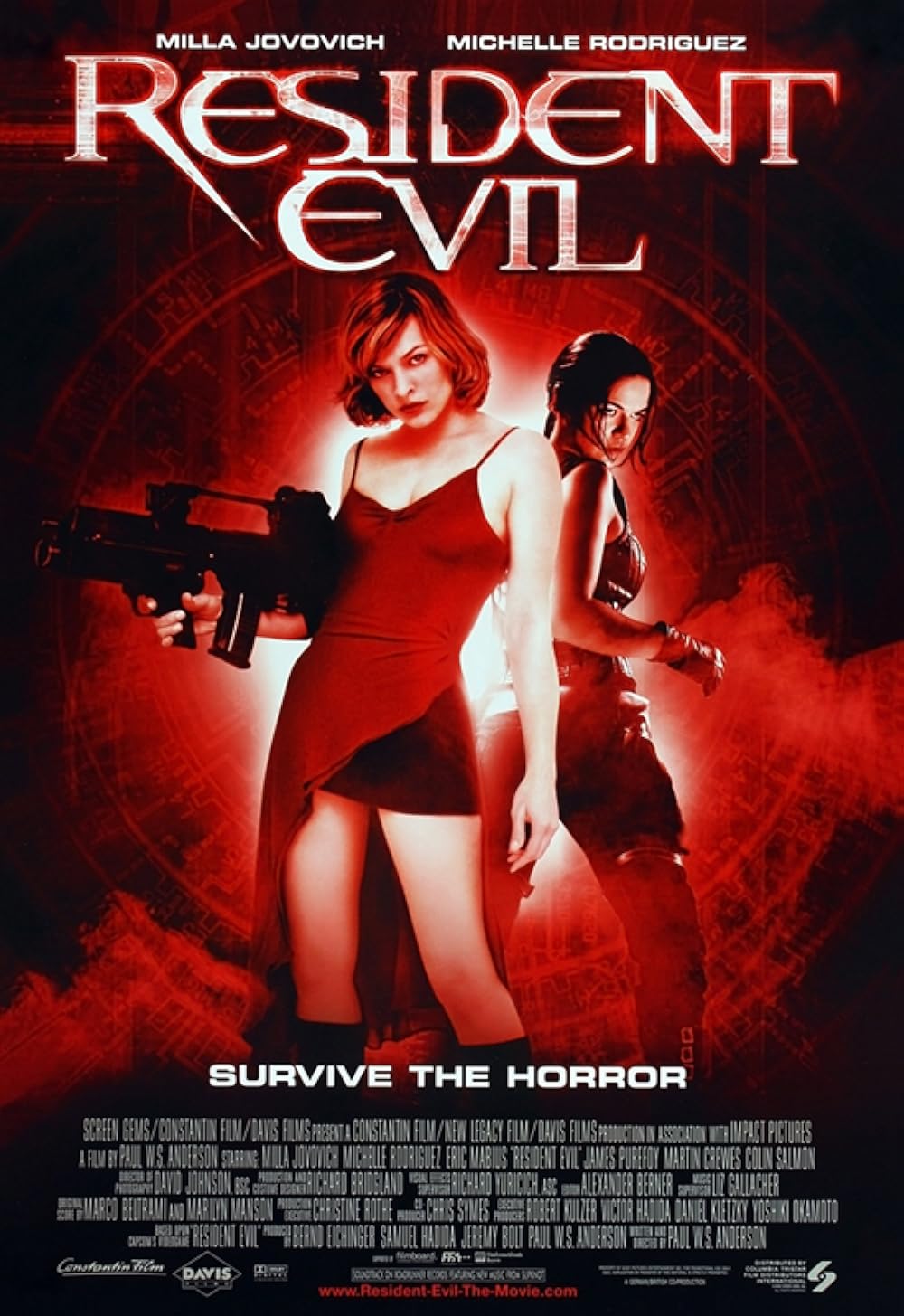} \\

\textcolor{blue}{{(a)}} horror,  & \textcolor{blue}{{(b)}} drama,  & \textcolor{blue}{{(c)}} action, & 
\textcolor{blue}{{(d)}} action, & \textcolor{blue}{{(e)}}  action,   \\
mystery, thriller & romance, \---  & sci-fi, \--- & adventure, fantasy & horror, sci-fi\\
\end{tabular}
\end{adjustbox}
\caption{Example of movie posters and their genres}
\label{fig:intro_challenge}
\end{figure}

Movie posters play a crucial role in shaping viewer expectations. They offer subtle clues about a film's genre, tone, and thematic elements, helping audiences decide whether a particular movie aligns with their interests and preferences \cite{intro_1}. For instance, Fig. \ref{fig:intro_challenge}.(a) a horror film poster showcases ominous imagery and dark colors, indicating to viewers that they can expect a fear experience, while \ref{fig:intro_challenge}.(b) a romantic poster utilizes bright colors and playful imagery to convey a cheerful and lighthearted tone. In online streaming platforms, users are bombarded with thumbnails and recommendations, and users scroll through these countless options \cite{intro_2}; movie posters have become an essential navigational tool since they can capture attention, arouse curiosity, and entice users to further exploration.

On the other hand, movie posters have a distinct advantage over trailers and plot summaries regarding their release timeline and visibility. While trailers and plot summaries typically become available closer to a film's release date, posters are often revealed well in advance, sometimes even months before the premiere. This early exposure allows posters to shape audience perceptions and generate anticipation long before other promotional materials are accessible. They also have the advantage of being widely used as thumbnails on OTT platforms, promotional materials, and social media, maximizing the movie's reach. 
In the literature, movie genre classification using trailer \cite{trailer_1} \cite{trailer_2} \cite{trailer_3} and plot summary \cite{plot_1} \cite{plot_2} \cite{plot_3} have been explored. 
However, research in movie genre classification using only posters has been relatively scarce \cite{poster_1} \cite{poster_2}. This gap \textbf{motivates} us to take on the genre classification task through movie posters.

Analyzing movie posters for genre identification presents a set of unique challenges. Many movie posters have a complex nature and limited information. For example, Fig. \ref{fig:intro_challenge}.(c) has intricate background, and Fig. \ref{fig:intro_challenge}.(d) has significantly less visual (and textual) information. Posters often represent multiple genres simultaneously, complicating classification tasks. Multi-label classification is required to accurately capture the diverse genre elements present in a single poster.
The analysis of movie posters for genre identification has traditionally focused on visual elements, overlooking the valuable information contained in textual details \cite{poster_3}. 
For example, Fig.\ref{fig:intro_challenge}.(e), the visual appearance of the poster suggests an action or sci-fi genre, characterized by dynamic imagery and futuristic elements. 
However, upon closer inspection of the textual details, the phrase \enquote{\footnotesize\emph{SURVIVE THE HORROR}} reveals the presence of a horror genre, adding a layer of genre complexity that may not be immediately apparent from visual cues alone. This highlights the significance of textual information in refining genre identification and ensuring accuracy in classification. \textbf{Inspired} by this, our study represents one of the earliest attempts to integrate visual and textual information from posters for movie genre identification.
Our major \textbf{contributions} to this paper are briefly mentioned below:
 
\emph{(i)} 
The first contribution of this work lies in its unique methodology for genre identification. Specifically, the approach involves the initial separation of textual content from visual elements within a poster, departing from the conventional treatment of the entire poster as a single image. This innovative use of bi-modal data significantly improves the model's ability to effectively discern genre information.
 
\emph{(ii)} 
Our approach involves the contrastive image and text pre-training that facilitates the extraction of visual and textual features, which are subsequently merged using bi-modal multi-head cross attention, followed by sequential multi-head self-attention. This unique fusion effectively captures both inter and intra-modal relationships, significantly enhancing the model's comprehension of the intricate complexities within genres.

\emph{(iii)} We performed extensive experiments and thoroughly evaluated our framework, comparing its performance against various contemporary architectures. Our findings demonstrate that our framework consistently outperformed major state-of-the-art methods, showcasing its superiority in multi-label genre identification tasks.
Moreover, the ablation study indicates the usefulness of each module of our framework.


The rest of the paper is structured as follows. 
The overview of the related literatures are presented in Section \ref{sec:related_work}.
The following Section \ref{3sec:method} describes the proposed methodology. 
After that, Section \ref{4sec:result} presents and analyzes the experimental results. 
Finally, Section \ref{5sec:conclusion} concludes this paper.
\section{Related Work}\label{sec:related_work}
\noindent 
Traditionally, researchers have explored a range of 
visual inputs (e.g., trailers \cite{trailer_3}, movie-clips \cite{clip_1}, facial frames \cite{trailer_2}) 
textual inputs (e.g., plot summaries \cite{plot_1}, screenplays \cite{plot_4}), and 
multimodal inputs integrating visual, textual, and audio data  \cite{multimodel_2}, \cite{multimodal_1}. 
In this paper, we concentrate on identifying multi-label movie genres solely from posters, 
which has been relatively limited in the existing literature \cite{poster_1}.

\vspace{.2cm}
\noindent 
\textbf{Visual Input:} 
Zhou et al. \cite{visual_1} extracted global visual descriptors from trailer keyframes and utilized KNN classifier for genre identification. 
In \cite{trailer_3} and \cite{trailer_4}, CNNs were employed to analyze trailers of LMTD (Labeled Movie Trailer Dataset) for detecting genres. 
A transformer network was used for genre identification from trailer clips in \cite{trailer_1}. 
In \cite{clip_1}, spatio-temporal features were extracted from video clips and hierarchical SVM was employed. 
Yadav et al. \cite{trailer_2} analyzed facial expressions in movie trailers to predict emotions and identify genres using an Inception-LSTM architecture.
Moving beyond the trailer, very few works have been found using movie poster images. 
Naïve Bayes classifier with global features was used in \cite{poster_3}. 
In \cite{poster_1}, CNN with GRAM layers was incorporated to capture style features from posters. 
YOLO and CNN models were engaged in identifying genres from detected objects of posters in \cite{poster_2}. 

\vspace{.2cm}
\noindent 
\textbf{Textual Input:} 
To analyze plot summaries for genre prediciton, BLSTM and GRU-based architectures were harnessed in \cite{plot_5} and \cite{plot_1}, respectively.
In \cite{plot_6} also, CNN with LSTMs and GRUs was employed for analyzing multi-lingual movie synopses.
CNN with the flow of emotions was used in \cite{plot_3} to capture emotions and genres from plot synopses. 
Wehrmann et al. \cite{plot_2} adopted a CNN architecture enriched with a self-attention mechanism for genre classification from synopses. 
Gorinski et al. \cite{plot_4} utilized movie screenplays for predicting various movie attributes, including genre, mood, etc. Their approach employed a multi-label encoder and LSTM-based decoder.

\vspace{.2cm}
\noindent 
\textbf{Multimodal Input:} 
%
Past studies integrated multiple data sources, such as poster images, synopsis texts, trailer videos, and audio, to enhance genre classification.
In \cite{ICLR_GMU} and \cite{CentralNet_ECCV}, the Gated Multimodal Unit (GMU) and CentralNet were employed, respectively, leveraging textual features from movie synopsis/metadata via Word2Vec and visual features from posters using CNN. 
GMU with transformer was utilized in \cite{multimodal_3} to integrate text, image, video, audio, and metadata features from diverse modalities. In \cite{multimodel_2}, fastText, fastVideo, VGG-19, and CRNN were employed to extract features from text (plot summary, metadata), trailers, posters, and audio, respectively. 
In \cite{multimodal_1}, various classifiers, including LSTM, KNN, SVM, MLP, and decision tree, were utilized to extract features from synopses, subtitles, trailers, posters, and audio.


\vspace{.2cm}
\noindent 
\textbf{Positioning of our work:} 
In the literature, there is a noticeable scarcity of research that focuses exclusively on movie posters for genre identification. 
The focus on other modalities has overshadowed the rich source of information contained within posters. Unlike trailers and clips, which provide dynamic visual sequences, posters offer a static yet highly curated representation of a film's aesthetic and thematic elements. Similarly, while textual summaries and screenplays provide comprehensive narrative details, posters distill this information into concise visual and textual elements that convey genre information at a glance. Furthermore, prior studies have only used visual information in the poster and neglected textual information. 
Our study is one of the earliest attempts to utilize both visual and textual information from movie posters for classifying multi-label genres.

\section{Proposed Methodology}
\label{3sec:method}

\begin{figure*}
\centering
\includegraphics[width=0.7\textwidth]{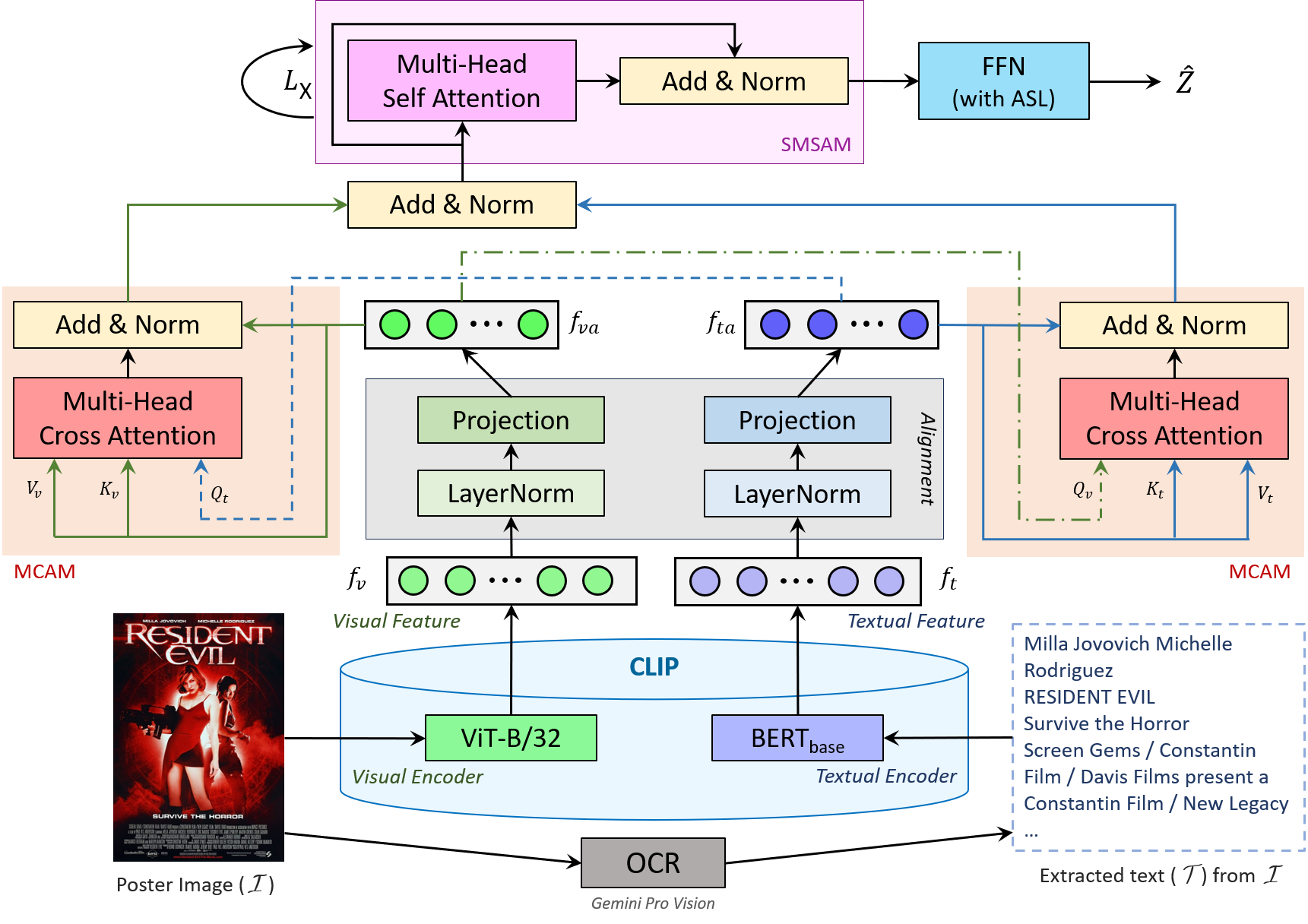}
\caption{Workflow of the proposed architecture}
\label{fig:workflow}
\end{figure*}

\noindent
In this section, we formulate our problem and then illustrate our proposed solution architecture.

\subsection{Problem Formulation}
\noindent
In this movie genre identification task, we are given a collection of $N$ movie poster images, represented as ${\cal{I}} = \{{\cal{I}}_1, {\cal{I}}_2, \ldots, {\cal{I}}_N\}$. Here, each ${\cal{I}}_i \in {\cal{I}}$ is a RGB image.
We are also given a set of genres denoted as $G = \{g_1, g_2, \ldots, g_M\}$, with $M$ indicating the total number of genres. Here, $M=13$. The $i^{th}$ movie poster image ${\cal{I}}_i \in {\cal{I}}$ is associated with a set of genres $G^{i} = \{g^{i}_1,g^{i}_2, \ldots, g^{i}_{k_i}\} \subseteq G$, where $k_i$ is the number of genres associated with ${\cal{I}}_i$. Recognizing that a single sample may belong to multiple positive classes, we strategically frame this problem into a multi-label classification task \cite{multilabel_classification}. Therefore, each ${\cal{I}}_i$ sample is labeled with a multi-label genre vector $Y^i \in \{0,1\}^{M}$.  
\begin{equation}
\label{eq:1}\small
{Y_j}^i=
\begin{cases}
    1~,  & ~~ \text{if $g_j \in G^{i}$} \\
    0~,  & ~~ \text{otherwise}
\end{cases}
\end{equation}
Our primary objective is to predict the correct multi-label genre vector $\hat{Y}^i$ for each ${\cal{I}}_i$. 

\subsection{Solution Architecture}
\label{sol_arch}
\noindent
Our proposed architecture begins by inputting a poster image, which undergoes an OCR (Optical Character Recognition) to extract text. Subsequently, both visual and textual features are extracted to undergo cross-modal understanding, followed by processing through a feed-forward neural network. 
The workflow of our solution architecture is pictorially shown in Fig. \ref{fig:workflow}, which we now present in detail.

\vspace{.2cm}
\noindent
\textbf{Encoding Movie Posters:} 
Our proposed framework involves the analysis of a movie poster image ${\cal{I}}_i$. 
The movie poster ${\cal{I}}_i$ is processed through Gemini-Pro-Vision \cite{gemini}-based OCR to extract text ${\cal{T}}_i = \left( t_{i1}, t_{i2}, \ldots, t_{ia}\right)$ which is tokenized into sub-word units, where $a$ is the number of tokens in the text ${\cal{T}}_i$.
For visual details, movie poster ${\cal{I}}_i$ divided it into distinct regions denoted as $r_i =\{ r_{i1},r_{i2}, \ldots, r_{ib} \};$ for $ r_{ij} \in \mathbb{R}^{b} $, where $b$ signifies the total count of regions.

The integration of language and image understanding at the semantic level is seamlessly achieved through the Contrastive Language-Image Pretraining (CLIP) network \cite{CLIP}. 
This architecture facilitates effective transfer learning by leveraging a contrastive learning approach. 
CLIP is extensively pre-trained on a vast dataset of 400 million image-text pairs from the internet, incorporating contrastive learning with both image and text exposure.
In alignment with these principles, our approach utilizes a pre-trained CLIP-based model \cite{CLIP} to extract features from both visual and textual details. 
    $\{{{f}_v}_i,{{f}_{t}}_i\} = \text{CLIP}({\cal{I}}_i,{\cal{T}}_i)$
where, ${{f}_v}_i \in \mathbb{R}^{D_i}$ and ${{f}_{t}}_i \in \mathbb{R}^{D_t}$ are the extracted visual and textaul feature, respectiviely. Here, $D_i$ and $D_t$  are the dimensions of visual and textual feature vectors, respectively.

\vspace{.2cm}
\noindent
\textbf{Cross Modal Understanding:} 
To capture the essence of integrating visual and textual features from movie poster ${\cal{I}}_i$ to create a cohesive multimodal representation, we implemented the cross-modal understanding module. This module comprises three crucial sub-modules:  
\emph{(i)} Alignment Module, 
\emph{(ii)} Multi-head Cross Attention Module ($MCAM$), and 
\emph{(iii)} Sequential Multi-head Self Attention Module ($SMSAM$), all of which are elaborated in detail below.

\emph{\textbf{(i) Alignment Module:}} 
Different modalities, such as text and images, capture information in diverse manners. Without proper alignment, the representations from these modalities lack comparability, resulting in inconsistent interpretations and hindering effective fusion \cite{alignment}. 
Feature alignment enhances the model's ability to leverage complementary information from different modalities. 
The dimensions of visual (${{f}_v}_i$) and textual feature (${{f}_{t}}_i$) may be different. 
We applied layer normalization \cite{Deep_learning} and linear projection functions to align these features in the same dimension $D_a$.
 \begin{equation}\small
    \label{eq:3}
     \begin{split}
                {{f}_{va}}_{i} = project(LN({{f}_v}_{i}))
       ~;~ {{f}_{ta}}_{i} = project(LN({{f}_{t}}_{i}))
    \end{split}   
\end{equation}
where, $project$ and $LN$ denote linear projection and layer normalization, respectively.

{\emph{\textbf{(ii) Multi-head Cross Attention Module (MCAM):}}}  
Cross attention proves to effectively exploit inter-feature relationships between visual and textual feature maps. Through a linear transformation, we derive $D_k$ dimensional \emph{queries} ($Q$) and \emph{keys} ($K$), and $D_v$ dimensional \emph{values} ($V$) for visual feature ${{f}_{va}}_i$, represented as 
$ \left( {{Q}_{v}}_i = {{f}_{va}}_i \cdot W_{vq}, {{K}_{v}}_i = {{f}_{va}}_i \cdot W_{vk}, {{V}_{v}}_i = {{f}_{va}}_i \cdot W_{vv} \right)$. 
Similarly, for the textual feature  ${{f}_{ta}}_i$, we obtain 
$\left( {{Q}_{t}}_i = {{f}_{ta}}_i \cdot W_{tq}, {{K}_{t}}_i = {{f}_{ta}}_i \cdot W_{tk}, {{V}_{t}}_i = {{f}_{ta}}_i \cdot W_{tv} \right)$, 
where, $\left\{ W_{vq}, W_{vk}, W_{tq}, W_{tk} \right\} \in \mathbb{R}^{{D_a} \times {D_k}}$ and 
$\left\{ W_{vv}, W_{tv} \right\} \in \mathbb{R}^{{D_a} \times {D_v}}$ are learnable weight matrices. 
Then, we applied multi-head cross attention $(MCA)$ to capture relationships and dependencies between different modalities. $MCA$ consists of multiple cross attention mechanisms in parallel \cite{attention}, where each attention head captures the essence of different aspects of inter-relationship between visual feature ${{f}_{va}}_i$ and ${{f}_{ta}}_i$. 
Cross attention utilizes scaled dot product attention $(SA)$ \cite{attention} to calculate the attention score. 
Here, we compute the visual (${{\cal{Z}}_{v}}_i$) and textual (${{\cal{Z}}_{t}}_i$) representations of cross attention as below:
%
\begin{equation}\scriptsize
    \label{eq:4}
        {{\cal{Z}}_{v}}_i  = \psi_m \left( \frac{{{Q}_{t}}_i \cdot \left( {{K}_{v}}_i\right)^{T}}{{\sqrt{D_k}}} \right) \cdot {{V}_{v}}_i ~;~ 
        {{\cal{Z}}_{t}}_i = \psi_m \left( \frac{{{Q}_{v}}_i \cdot \left( {{K}_{t}}_i\right)^{T}} {{\sqrt{D_k}}} \right) \cdot {{V}_{t}}_i
\end{equation}
where, $\psi_m$ denotes \emph{softmax} activation \cite{Deep_learning}, and ${\mathcal{X}}^{T}$ represents transpose of matrix $\mathcal{X}$.
\noindent
The multi-head cross attention is empowered with multiple heads. The concurrent cross attention is collectively computed as follows:
\begin{equation}\small
\label{eq:5}
\begin{split}
    {{{\cal{Z}}_{v}}^{cc}_i} = [ {{{\cal{Z}}_{v}}^{1}_i}, {{{\cal{Z}}_{v}}^{2}_i},\ldots, {{{\cal{Z}}_{v}}^{h}_i}] 
    ~;~{{{\cal{Z}}_{t}}^{cc}_i} = [ {{{\cal{Z}}_{t}}^{1}_i}, {{{\cal{Z}}_{t}}^{2}_i},\ldots, {{{\cal{Z}}_{t}}^{h}_i}]
\end{split}
\end{equation}
where, $h$ signifies the number of attention heads; ${{{\cal{Z}}_{v}}^{cc}_i}$ and ${{{\cal{Z}}_{t}}^{cc}_i}$ are the concatenated visual and textual representations, respectively. 
We observed information loss after applying multi-head cross attention. 
To address this issue, we employed residual connection, which eventually improved the performance. 
After that, we fused the visual and textual representations by leveraging layer normalization ($LN$).
%
\begin{equation}\small
\centering
    \label{eq:6}
    \begin{split}
        {{\cal{Z}}{{f}_{v}}_i} = LN \left( {{{\cal{Z}}_{v}}^{cc}_i} + {{f}_{va}}_i\right) ~;~
        {{\cal{Z}}{{f}_{t}}_i} = LN \left( {{{\cal{Z}}_{t}}^{cc}_i} + {{f}_{ta}}_i\right) ~;\\
    {f_{vt}}_i = LN \left( {{\cal{Z}}{{f}_{v}}_i} +  {{\cal{Z}}{{f}_{t}}_i}\right)
    \end{split}
\end{equation}

\emph{\textbf{(iii) Sequential Multi-head Self Attention Module (SMSAM):}} 
Multi-head Self Attention ($MSA$) can directly learn long-range dependencies and intra-relationships in input features. $MSA$ also contains multiple parallel attention layers. Each head uses SA \cite{attention}, wherein the input consists of queries and keys with dimensionality of $D_k$, and values with dimensionality of $D_v$. $MSA$ is computed as follows: 
\begin{equation}\small
\label{eq:8}
\begin{split}
    SA(Q,K,V) = \psi_m \left( {QK^T}\diagup{\sqrt{D_k}} \right)V ;\\
    MSA \left(Q,K,V\right) = [head^1, head^2, \ldots, head^h] ;\\
    head^i = SA(Q\cdot W^i_q,~ K\cdot W^i_k,~ V\cdot W^i_v)
\end{split}
\end{equation}

\noindent
where, a set of queries, keys, and values are simultaneously packed together and form corresponding $Q$, $K$, $V$ matrices. $W^i_q \in \mathbb{R}^{D_{vt} \times D_k}$, $W^i_k \in \mathbb{R}^{D_{vt} \times D_k}$ and $W^i_v \in \mathbb{R}^{D_{vt} \times D_v}$ are learnable metrices. $D_{vt}$ is the dimension of ${{f}_{vt}}_i$.

We applied this $MSA$ in sequential connection to a rich representation. It can be formulated as below:
\begin{equation}\small
    \label{eq:10}
    z_0 = {{f}_{vt}}_i ~; ~ 
    z_l = MSA \left( LN \left( z_{l-1}\right)  \right) + z_{l-1} ~ ; 
\end{equation}
where, $l = 1,2,\ldots,L$. Here $L=4$ is the number of $MSA$ used sequentially. After applying multiple $MSA$, we get the final fused representation: $y = LN \left( z_L \right)$.

\vspace{.2cm}
\noindent
\textbf{Feed Forward Neural Network (FFN):}  
The final stage of our model incorporates an FFN comprising two hidden layers featuring 512 and 128 nodes, respectively, which uses the ReLU (Rectified Linear Unit) activation function \cite{Deep_learning}. The output layer contains $M$ nodes and employs the sigmoid output function \cite{Deep_learning}. 
Conclusively, when presented with a movie poster image ${\cal{I}}_i$, our framework provides a genre confidence score vector $\hat{Z}^i = \left(  \hat{Z}^i_1, \hat{Z}^i_2,\ldots, \hat{Z}^i_M \right)$.

\vspace{.2cm}
\noindent
\textbf{Loss Function:}  
Multi-label datasets are mostly imbalanced and have more negative samples than positive ones for a particular class. We use Asymmetric Loss $(ASL)$ to address this issue \cite{asl_cvpr}. The main objective of $ASL$ is to provide control over positive and negative sample imbalance in the optimization of multi-label classification. 
It prioritizes learning of more challenging samples, particularly those belonging to minority classes in the given datasets. $ASL$ achieves this by adjusting specific weights to each class in the loss function, considering the complexity of the classification task. 
Consequently, the loss function assigns higher weights to minority class samples, which involve greater classification complexity, while allocating lower weights to majority class samples that are comparatively simpler to classify.
Given the input movie poster image ${\cal{I}}_i$, we can predict its genre confidence score vector $\hat{Z}^i = \left(  \hat{Z}^i_1, \hat{Z}^i_2,\ldots, \hat{Z}^i_M \right)$. using our framework. We employ the asymmetric loss to calculate the loss as below:
\begin{equation}\label{eq:11}
\scriptsize 
\begin{split}
     {\cal{L}}_{ASL} = \frac{1}{M}\sum_{j=1}^{M} \left({\cal{L}}_j^+ +  {\cal{L}}_j^-\right);
     \quad {\cal{L}}_j^+ = 
        Y^i_j \left(1-\hat{Z}^i_j \right)^{\gamma ^{+}} \log{\left( \hat{Z}^i_j \right)};\\
     {\cal{L}}_j^- = 
        (1 - Y^i_j)
        P_{\epsilon} \left( \hat{Z}^i_j \right)^{\gamma ^{-}} \log{ \left( 1-P_{\epsilon} \left( \hat{Z}^i_j \right) \right)}
\end{split}
\end{equation}
where, $ Y^i_j$ denotes the ground truth for $j^{th}$ genre $g_j \in G$ for image ${\cal{I}}_i$ calculated through Eqn. \ref{eq:1}. $P_{\epsilon} \left( \hat{Z}^i_j \right) $ is shifted probability of $\hat{Z}^i_j$. Shifted probability is the probability assigned to negative samples after applying hard thresholding using the margin $\epsilon$ for adjustment. It is defined as below:

\begin{equation} \label{eq:eq12}\small
    P_{\epsilon} \left( \hat{Z}^i_j \right) = \max \left( \hat{Z}^i_j - \epsilon,~0 \right)
\end{equation}
The overall cost is calculated by taking an average of losses across all samples in the training dataset. 
Empirically, we set $\gamma ^{+} = 3$, $\gamma ^{-}=4$, $\epsilon=0.2$ for our experiments.

\vspace{.2cm}
\noindent
\textbf{Post-processing:}  
%
In this step, we apply quantization on the confidence score vector $\hat{Z}^i$ to get the final output vector $\hat{Y^i} \in \{0,1\}^M$, as below:
%
\begin{equation}
\label{eq:13}\small
\hat{Y_j}^i=
\begin{cases}
    1~,  & ~~ \text{if $ \hat{Z}^i_j > \uptau $} \\
    0~,  & ~~ \text{otherwise}
\end{cases}
\end{equation}
where, we have empirically chosen the threshold $\uptau$ to be 0.5. 

\section{Experiments and Discussions}
\label{4sec:result}

\noindent
In this section, we discuss experiments performed to check the efficacy of our proposed model. We begin by presenting the database employed to perform the experiments. 

\subsection{Database Employed}
\label{subsec:data_emp}
\begin{figure}[!h]
    \centering
    \includegraphics[width=0.9\linewidth]{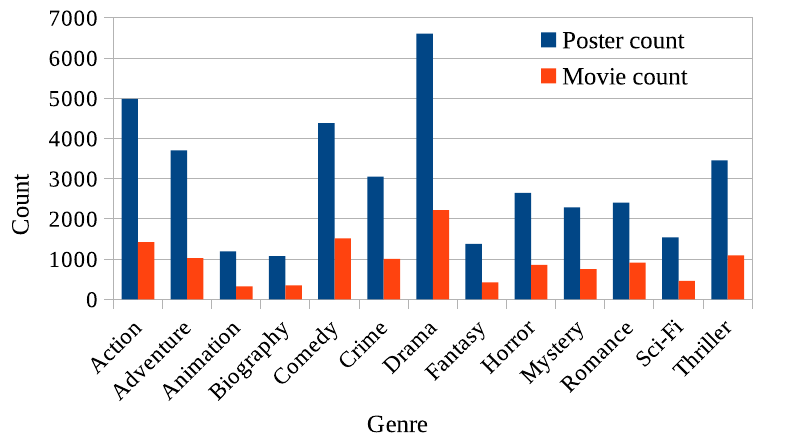}
    \caption{Dataset details: movie and poster count across genres}
    \label{fig:dataset}
\end{figure}

\noindent
Our framework undergoes evaluation using the IMDb dataset, a collection of authentic movie poster images coupled with corresponding genres sourced from the Internet Movie Database (IMDb: \href{https://developer.imdb.com/non-commercial-datasets}{\emph{https://developer.imdb.com/non-commercial-datasets}}). This dataset serves as a multi-label movie genre classification dataset. It comprises 4464 films, each associated with 1 to 5 posters. Each movie poster is linked to 1 to 3 genres, resulting in a total of 13882 unique movie posters. In total, there are 13 distinct genres, including 
action, 
adventure, 
animation, 
biography, 
comedy, 
crime, 
drama, 
fantasy, 
horror, 
mystery, 
romance, 
sci-fi, and 
thriller. In Fig. \ref{fig:dataset}, we present the detailed genre-wise dataset distribution. 
We perform a random split of the dataset into a training set ${\cal{DB}}_{train}$, validation set ${\cal{DB}}_{valid}$ and testing set ${\cal{DB}}_{test}$ at the ratio of 0.8, 0.1 and 0.1, respectively.
As a matter of fact, we leverage ${\cal{DB}}_{train}$ for model training.

\subsection{Results}
\noindent
In this subsection, we start with experimental settings and evaluation metrics, followed by a comparative analysis, ablation study, and various additional evaluations.

\vspace{.2cm}
\noindent
\textbf{Experimental Settings:}   
Our experimentation was conducted using the TensorFlow 2.8 framework, running on Python 3.9.13, on an Ubuntu 20.04.2 LTS-based machine. The machine specifications include an AMD EPYC 7552 Processor operating at 2.20 GHz with 48 CPU cores and 256 GB RAM equipped with a 40 GB NVIDIA A100-PCIE GPU.

Throughout this paper, all presented results were obtained from the testing set ${\cal{DB}}_{test}$. Model hyperparameters were tuned and set during training, with a focus on optimizing performance over the validation set ${\cal{DB}}_{valid}$. 
All models are trained for epoch 100. The Adam optimizer parameters were chosen as follows: 
initial learning rate = $10^{-4}$; 
exponential decay rates for 1$^{st}$ and 2$^{nd}$ moment estimates, i.e., $\beta1=0.9$, $\beta2=0.999$; and
zero-denominator removal parameter ($\upvarepsilon$) = $10^{-8}$.
For the early stopping strategy, we set the patience parameter to 10 epochs, and fixed mini-batch size of 32.

\vspace{.2cm}
\noindent
\textbf{Evaluation Metrics:}  
For the overall model performance evaluation, we report the 
macro F1 (${\cal{F}}_{m}$) \%, macro balanced accuracy (${\cal{BA}}_{m}$) \%, 
micro F1 (${\cal{F}}_{\mu}$) \%, micro balanced accuracy (${\cal{BA}}_{\mu}$) \%, 
weighted F1 (${\cal{F}}_{w}$) \%, weighted balanced accuracy (${\cal{BA}}_{w}$) \%, 
samples F1 (${\cal{F}}_{s}$) \%, samples balanced accuracy (${\cal{BA}}_{s}$) \%, 
hamming loss (${\cal{HL}}$), and hit ratio of first genre (${\cal{H}}it$) as our metrics \cite{ICLR_GMU,eval_metrics_acm}. 
For genre-wise analysis, we use precision (${\cal{P}}$) \%, recall (${\cal{R}}$) \%, F1-score (${\cal{F}}$) \%, balanced accuracy (${\cal{BA}}$) \%, and specificity (${\cal{S}}p$) \% \cite{eval_metrics_acm}.

\begin{table*}[!t]
\centering
\caption{Comparison with Baseline and SOTA methods on bi-modality (image and text) 
}
\begin{adjustbox}{width=.7\textwidth} 
\begin{tabular}{c|l|c|c|c|c|c|c|c|c|c}
\cline{2-11}
\multicolumn{1}{l}{} & \textbf{Methods (bi-modal)} &  
${\cal{F}}_{m}$ & ${\cal{BA}}_{m}$ & ${\cal{F}}_{\mu}$ & ${\cal{BA}}_{\mu}$ & ${\cal{F}}_{w}$ & ${\cal{BA}}_{w}$ & ${\cal{F}}_{s}$ & ${\cal{BA}}_{s}$ & ${\cal{HL}}$ \\
\hline
\hline
 & \begin{tabular}[c]{@{}l@{}}MobileNetV2 $+$  W2V \end{tabular} & 
 31.53 & 61.34 & 47.49 & 65.96 & 38.87 & 62.90 & 44.39 & 66.00 & 0.1839 \\ \cline{2-11}
 & \begin{tabular}[c]{@{}l@{}}VGG19 $+$  W2V \end{tabular} & 
 39.50 & 63.08 & 51.69 & 68.21 & 48.23 & 65.75 & 48.51 & 68.31 & 0.1726 \\ \cline{2-11}
 
 & \begin{tabular}[c]{@{}l@{}}InceptionV3 $+$  W2V \end{tabular} & 
 36.09 & 61.92 & 48.99 & 66.73 & 45.65 & 64.77 & 46.43 & 66.94 & 0.1786 \\ \cline{2-11}
 
 & \begin{tabular}[c]{@{}l@{}}ResNet50V2 $+$  W2V \end{tabular} & 
 39.20 & 63.68 & 52.35 & 68.47 & 48.30 & 66.46 & 49.36 & 68.68 & 0.1673 \\ \cline{2-11}
 
 & \begin{tabular}[c]{@{}l@{}}DenseNet121 $+$  W2V \end{tabular} & 
 45.55 & 66.60 & 55.68 & 70.58 & 53.54 & 69.09 & 52.91 & 70.80 & 0.1651 \\ \cline{2-11}
 
 & \begin{tabular}[c]{@{}l@{}}EfficientNetV2B2 $+$  W2V \end{tabular} & 
 44.42 & 66.76 & 58.45 & 72.14 & 53.78 & 69.68 & 55.82 & 72.23 & 0.1557 \\ \cline{2-11}
 
 & \begin{tabular}[c]{@{}l@{}}MobileNetV2 $||$  W2V \end{tabular} & 
 22.75 & 57.74 & 39.91 & 62.13 & 29.60 & 59.20 & 36.93 & 62.27 & 0.2021 \\ \cline{2-11}
 
 & \begin{tabular}[c]{@{}l@{}}VGG19 $||$  W2V \end{tabular} & 
 35.53 & 62.13 & 48.97 & 66.57 & 44.28 & 64.79 & 44.90 & 66.73 & 0.1695 \\ \cline{2-11}
 
 & \begin{tabular}[c]{@{}l@{}}InceptionV3 $||$  W2V \end{tabular} & 
 39.57 & 63.99 & 50.84 & 67.71 & 47.75 & 66.20 & 47.28 & 67.93 & 0.1736 \\ \cline{2-11}
 
 & \begin{tabular}[c]{@{}l@{}}ResNet50V2 $||$  W2V \end{tabular} & 
 40.12 & 64.34 & 52.78 & 68.77 & 48.84 & 66.78 & 49.53 & 68.87 & 0.1682 \\ \cline{2-11}
 
 & \begin{tabular}[c]{@{}l@{}}DenseNet121 $||$  W2V \end{tabular} & 
 43.04 & 65.87 & 55.12 & 70.03 & 51.70 & 68.63 & 52.25 & 70.31 & 0.1612 \\ \cline{2-11}

 & \begin{tabular}[c]{@{}l@{}}EfficientNetV2B2 $||$  W2V \end{tabular} & 
 46.01 & 67.24 & 58.51 & 72.22 & 54.81 & 69.96 & 56.07 & 72.45 & 0.1565 \\ \cline{2-11}

 & BLIP \cite{BLIP} & 
 64.11 & 76.18 & \textbf{69.99} & \textbf{79.58} & 68.68 & 77.88 & \textbf{68.13} & \textbf{79.83} & \textbf{0.1208} \\ \cline{2-11}

\multirow{-15}{*}{\rotatebox{90}{Baseline}} & CLIP \cite{CLIP}& 
\textbf{64.81} & \textbf{76.70} & 69.67 & 79.29 & \textbf{69.07} & \textbf{78.18} & 67.90 & 79.68 & 0.1214 \\ \cline{2-11}

& \multicolumn{1}{l|}{\cellcolor[HTML]{EFEFEF}Our improvement} & 
\cellcolor[HTML]{EFEFEF}3.42 & \cellcolor[HTML]{EFEFEF}3.09 & \cellcolor[HTML]{EFEFEF}2.35 & \cellcolor[HTML]{EFEFEF}2.80 & \cellcolor[HTML]{EFEFEF}3.02 & \cellcolor[HTML]{EFEFEF}2.52 & \cellcolor[HTML]{EFEFEF}3.45 & \cellcolor[HTML]{EFEFEF}2.90 & \cellcolor[HTML]{EFEFEF}0.26\% \\
\hline \hline

 & GMU \cite{ICLR_GMU} & 
 51.98 & 69.51 & 61.26 & 73.71 & 58.96 & 72.20 & 57.58 & 73.85 & 0.1455 \\ \cline{2-11}

 & CentralNet \cite{CentralNet_ECCV} & 
\textbf{53.59} & \textbf{70.72} & \textbf{62.67} & \textbf{74.64} & \textbf{60.30} & \textbf{73.18} & \textbf{59.92} & \textbf{74.81} & \textbf{0.1423} \\ \cline{2-11}

\multirow{-3}{*}{\rotatebox{90}{SOTA}} &\multicolumn{1}{l|}{\cellcolor[HTML]{EFEFEF}Our improvement} & 
\cellcolor[HTML]{EFEFEF}14.63 & \cellcolor[HTML]{EFEFEF}9.07 & \cellcolor[HTML]{EFEFEF}9.67 & \cellcolor[HTML]{EFEFEF}7.73 & \cellcolor[HTML]{EFEFEF}11.79 & \cellcolor[HTML]{EFEFEF}7.52 & \cellcolor[HTML]{EFEFEF}11.65 & \cellcolor[HTML]{EFEFEF}7.95 & \cellcolor[HTML]{EFEFEF}15.30\% \\ 
\hline \hline
\multicolumn{1}{l}{}& \textbf{Ours} & 
\textbf{68.23} & \textbf{79.79} & \textbf{72.34} & \textbf{82.38} & \textbf{72.09} & \textbf{80.69} & \textbf{71.58} & \textbf{82.73} & \textbf{0.1205} \\
\cline{2-11} 
\multicolumn{11}{r}{\enquote{$||$} and \enquote{$+$} denotes \emph{concatenation} and \emph{linear addition}, respectively, used for fusion.}
\end{tabular}
\end{adjustbox}
\label{tab:comparison}
\end{table*}

\vspace{.2cm}
\noindent
\textbf{Comparative Study:}
We devised a comprehensive comparison by implementing various Baseline and State-of-the-Art (SOTA) methods.  For our baselines, we leveraged VGG19 \cite{VGG19_ICLR}, ResNet50V2 \cite{ResNet50V2_ECCV}, DenseNet121 \cite{DenseNet121_CVPR}, MobileNetV2 \cite{MobileNetV2_CVPR}, InceptionV3 \cite{InceptionV3_CVPR}, and EfficientNetV2B2 \cite{EfficientNetV2B2_ICML} to extract visual features. We utilized Google Word2Vec (W2V) \cite{Word2Vec_1} \cite{Word2Vec_2} with Long Short Term Memory (LSTM) \cite{LSTM} to extract textual features.  These features were fused using either concatenation or linear addition techniques and subsequently fed into a Feedforward Network (FFN). 
Additionally, we implemented BLIP \cite{BLIP} and CLIP \cite{CLIP} to extract visual and textual features. 
We then fused these features using linear addition and fed them into FFN.
%
Furthermore, we implemented GMU \cite{ICLR_GMU} and CentralNet \cite{CentralNet_ECCV} as SOTA. It was originally designed to utilize movie posters and summaries to predict genres; we modified these methods to operate on movie posters and extracted text from posters. 
This adjustment allowed for a more direct comparison with our framework. 

Table \ref{tab:comparison} presents a comprehensive performance comparison between our proposed framework, baseline and SOTA methods. Notably, our framework demonstrates superior performance across all evaluation metrics compared to the baseline and SOTA methods. 
In the baseline comparison, transformer-based models,  BLIP \cite{BLIP} and CLIP \cite{CLIP} significantly outperform traditional deep learning architectures. However, our framework surpasses even these transformer-based models across all evaluation metrics, showcasing its robustness and effectiveness. 
Compared to SOTA methods, our framework exhibits substantial improvements, surpassing them by at least 7.5\% across all evaluation metrics. This notable performance enhancement can be attributed to incorporating two key modules, $SMSAM$ and $MCAM$, which effectively enrich intra-feature and inter-modal relations, thereby enhancing the overall performance of our framework.

\begin{table}[!t]
\centering
\caption{Ablation study on modality}
\begin{adjustbox}{width=0.49\textwidth} 
\begin{tabular}{c|l|c|c|c|c|c|c|c|c|c}
\hline
\textbf{Modality} & \textbf{Methods} &  
${\cal{F}}_{m}$ & ${\cal{BA}}_{m}$ & ${\cal{F}}_{\mu}$ & ${\cal{BA}}_{\mu}$ & ${\cal{F}}_{w}$ & ${\cal{BA}}_{w}$ & ${\cal{F}}_{s}$ & ${\cal{BA}}_{s}$ & ${\cal{HL}}$ \\ \hline\hline
Text & CLIP \cite{CLIP} & 
\textbf{4.71} & \textbf{50.00} & \textbf{23.09} & \textbf{55.07} & \textbf{9.57} & \textbf{50.00} & \textbf{22.90} & \textbf{55.05} & \textbf{0.2263} \\
\hline
\rowcolor[HTML]{EFEFEF} 
\multicolumn{2}{c|}{\cellcolor[HTML]{EFEFEF} Our improvement} & 
63.51 & 29.79 & 49.25 & 27.30 & 62.52 & 30.69 & 48.68 & 27.68 & 46.74\% \\ \hline \hline
 & MobileNetV2 \cite{MobileNetV2_CVPR} & 
 34.04 & 62.48 & 48.04 & 66.27 & 41.60 & 64.23 & 44.33 & 66.41 & 0.1840 \\  \cline{2-11}
 & ResNet50V2 \cite{ResNet50V2_ECCV} & 
 35.32 & 61.58 & 48.56 & 66.46 & 44.27 & 64.08 & 44.64 & 66.46 & 0.1770 \\ \cline{2-11} 
 & VGG19 \cite{VGG19_ICLR} & 
 35.81 & 62.42 & 50.58 & 67.55 & 45.08 & 65.14 & 47.32 & 67.66 & 0.1730 \\ \cline{2-11} 
 & InceptionV3 \cite{InceptionV3_CVPR}& 
 38.59 & 63.27 & 48.49 & 66.39 & 45.53 & 65.12 & 44.67 & 66.65 & 0.1743 \\ \cline{2-11}
 & DenseNet121 \cite{DenseNet121_CVPR} & 
 42.99 & 64.96 & 55.13 & 70.16 & 51.95 & 67.79 & 52.09 & 70.29 & 0.1642 \\ \cline{2-11}
 & EfficientNetV2B2 \cite{EfficientNetV2B2_ICML} & 
 46.95 & 66.88 & 57.03 & 71.31 & 54.63 & 69.31 & 53.95 & 71.48 & 0.1600 \\ \cline{2-11}
 & BLIP \cite{BLIP} & 
 61.63 & 74.95 & 69.07 & 78.63 & 67.07 & 77.21 & 66.90 & 78.94 & 0.1212 \\
 \cline{2-11}
 \multirow{-8}{*}{Image} & CLIP \cite{CLIP} & 
\textbf{64.58} & \textbf{76.48} & \textbf{70.19} & \textbf{79.77} & \textbf{69.19} & \textbf{78.25} & \textbf{68.51} & \textbf{80.04} & \textbf{0.1206} \\
\hline
\rowcolor[HTML]{EFEFEF} 
\multicolumn{2}{c|}{\cellcolor[HTML]{EFEFEF}Our improvement} & 
3.65 & 3.31 & 2.15 & 2.60 & 2.90 & 2.44 & 3.07 & 2.69 & 0.04\% \\
\hline \hline
Image + Text & \textbf{Ours} & 
\textbf{68.23} & \textbf{79.79} & \textbf{72.34} & \textbf{82.38} & \textbf{72.09} & \textbf{80.69} & \textbf{71.58} & \textbf{82.73} & \textbf{0.1205} \\
\hline
\end{tabular}
\end{adjustbox}
\label{tab:feature_ablation}
\end{table}

\vspace{.2cm}
\noindent
\textbf{Modality Ablation Study:} 
Table \ref{tab:feature_ablation} illustrates the performance achieved with different modalities: text, image, and their combination. As our framework's core module, $MCAM$, requires two modalities to operate,  we did not conduct experiments using single modalities (text or image) with it.

In the text modality, methods displayed subpar performance, mainly due to the limited textual content on most movie posters, which directly describes a genre of the movie without the context of visual features. This text typically consists of cast/crew names and movie titles.
In the image modality, transformer-based models exhibited superior performance compared to conventional deep learning architectures. Notably, CLIP \cite{CLIP} emerged as the top-performing model in this modality.
However, in the image + text modality, our framework demonstrated the best results, surpassing all models exploring a single modality.

We have 12 combinations of contemporary baseline deep architectures, and 2 transformer-based baseline methods (i.e., BLIP \cite{BLIP} and CLIP \cite{CLIP}), in Table \ref{tab:comparison}. 
From Table \ref{tab:comparison} and \ref{tab:feature_ablation},  we have the following major observations:

\emph{\textbf{(i)}} Our framework outperformed all other methods, compared here.
    
\emph{\textbf{(ii)}} Integrating the text modality with the image significantly enhanced performances across all evaluation metrics for four baseline method combinations \cite{ResNet50V2_ECCV,VGG19_ICLR,InceptionV3_CVPR}, and BLIP \cite{BLIP}.

\emph{\textbf{(iii)}} The performance of seven contemporary baseline combinations \cite{VGG19_ICLR, DenseNet121_CVPR, MobileNetV2_CVPR, InceptionV3_CVPR, EfficientNetV2B2_ICML}, and CLIP \cite{CLIP} slightly increased in some evaluation metrics after using text information as supplementary.

\emph{\textbf{(iv)}} The performance of one baseline combination (i.e., MobileNetV2 || W2V) \cite{MobileNetV2_CVPR} exhibited a slight degradation across all evaluation metrics, indicating that the text modality did not effectively complement the image modality in this case.


From these observations, it is evident that leveraging textual information from movie posters as supplementary features can significantly enhance the accuracy of movie genre identification using posters.

\vspace{.2cm}
\noindent
\textbf{Model Ablation Study:}
Table \ref{tab:model_ablation} provides an ablation study of our proposed framework. Removing the $MCAM$ module results in a performance loss across all evaluation metrics due to the absence of inter-modality feature relations. Similarly, excluding the $SMSAM$ module leads to performance degradation in all evaluation metrics, as it fails to capture intra-modality feature relations.
Furthermore, when both the $MCAM$ and $SMSAM$ modules are discarded, we observe a decline in performance by at least 3\%, i.e., equivalent to CLIP. This suggests that our framework relies on reliable attention allocation to various modality features and possesses a better-discriminating capability of fused features, enhancing our model's effectiveness.

\begin{table}[!t]
\centering
\caption{Model Ablation study on each module}
\begin{adjustbox}{width=0.49\textwidth} 
\begin{tabular}{l|c|c|c|c|c|c|c|c|c}
\hline
\multicolumn{1}{l|}{\bf Model}  &  
${\cal{F}}_{m}$ & ${\cal{BA}}_{m}$ & ${\cal{F}}_{\mu}$ & ${\cal{BA}}_{\mu}$ & ${\cal{F}}_{w}$ & ${\cal{BA}}_{w}$ & ${\cal{F}}_{s}$ & ${\cal{BA}}_{s}$ & ${\cal{HL}}$ \\
\hline\hline
Ours ($L=4$) & \textbf{68.23} & \textbf{79.79} & \textbf{72.34} & \textbf{82.38} & \textbf{72.09} & \textbf{80.69} & \textbf{71.58} & \textbf{82.73} & \textbf{0.1205} \\
\hline
Ours ($L=1$) & 67.89 & 79.47 & 71.73 & 82.10 & 71.65 & 80.49 & 71.05 & 82.38 & 0.1239 \\
\hline
Ours $- SMSAM$ & 67.77 & 79.50 & 71.89 & 82.01 & 71.51 & 80.43 & 71.20 & 82.32 & 0.1220 \\
\hline
Ours ($L=4$) $- MCAM$ & 67.07 & 79.32 & 71.64 & 82.11 & 71.10 & 80.36 & 71.04 & 82.42 & 0.1248 \\
\hline
Ours ($L=1$) $- MCAM$  & 66.77 & 78.77 & 71.31 & 81.82 & 70.93 & 80.10 & 70.59 & 82.04 & 0.1257 \\
\hline
Ours $- MCAM - SMSAM$ & 64.81 & 76.70 & 69.67 & 79.29 & 69.07 & 78.18 & 67.90 & 79.68 & 0.1214 \\
\hline
\multicolumn{10}{r}{"Ours ($L=4$)": Our overall method with $4$ sequential $MSA$; 
"$-~{\mathcal{M}}$": Ablating $\mathcal{M}$ module from ours.}
\end{tabular}
\end{adjustbox}
\label{tab:model_ablation}
\end{table}





\vspace{.2cm}
\noindent
\textbf{Genre-wise Analysis:} 
Table \ref{tab:genrewise} demonstrated the genre-wise performance of our framework. Notably, our framework excelled in the animation genre, achieving the highest scores in ${\cal{BA}}$ (balanced accuracy), 
${\cal{F}}$ (F1-score), 
${\cal{P}}$ (precision), 
${\cal{R}}$ (recall) and 
${\cal{S}}p$ (specificity). 
Additionally, it performed impressively in the adventure, comedy, action, romance, and horror genres by attaining more than 80\% ${\cal{BA}}$. 
%
However, it encountered challenges and performed average in 
biography, mystery, fantasy, and sci-fi genres with respect to ${\cal{F}}$; 
however, these genres attained high ${\cal{S}}p$, i.e., 95.37\%, 93.97\%, 94.15\%, and 95.86\%, respectively, 
indicating their proficiency in correctly classifying negative samples (high true negative), and low ${\cal{F}}$ indicating low true positive, which is caused by multi-label classification \cite{multilabel_classification}.



\begin{table*}
\centering
\caption{Genre-wise performance of our model}
\begin{adjustbox}{width=0.98\textwidth} 

\begin{tabular}{c|c|c|c|c|c|c|c|c|c|c|c|c|c}
\cline{2-14}
\cline{2-14}
 \multicolumn{1}{c}{} & \multicolumn{1}{c|}{Animation}  & Adventure & Comedy & Action & Romance & Horror & Drama & Sci-Fi & Biography & Crime & Thriller & Mystery & Fantasy \\
 \hline\hline
 ${\cal{BA}}$ & \textbf{94.74} & 86.71 & 85.35 & 85.26 & 84.91 & 80.21 & 79.25 & 78.72 & 76.77 & 74.35 & 73.51 & 70.33 & 67.20 \\
\hline
${\cal{F}}$ & \textbf{91.29} & 82.42 & 78.76 & 83.11 & 69.33 & 72.05 & 77.20 & 65.00 & 54.01 & 58.94 & 60.55 & 51.96 & 42.32 \\
\hline
${\cal{P}}$ & \textbf{92.12} & 82.59 & 78.01 & 79.02 & 63.76 & 84.18 & 74.75 & 68.82 & 50.39 & 57.73 & 61.56 & 58.56 & 44.60 \\
\hline
${\cal{R}}$ & \textbf{90.48} & 82.26 & 79.52 & 87.64 & 75.96 & 62.98 & 79.82 & 61.58 & 58.18 & 60.20 & 59.57 & 46.70 & 40.26 \\
\hline
${\cal{S}}p$ & \textbf{99.00} & 91.17 & 91.18 & 82.88 & 93.86 & 97.43 & 78.68 & 95.86 & 95.37 & 88.51 & 87.44 & 93.97 & 94.15 \\
\hline
\end{tabular}
\end{adjustbox}
\label{tab:genrewise}
\end{table*}

\begin{figure*}
    \centering
\includegraphics[width=0.9\linewidth]{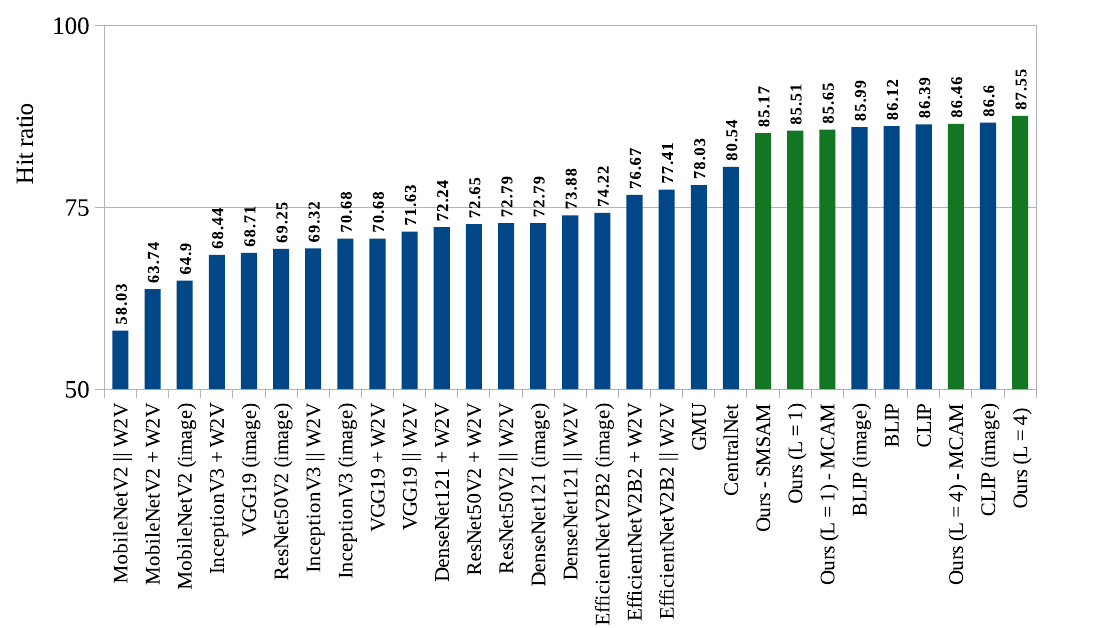}
\caption{Performance analysis on hit ratio $({\cal{H}}it)$.  
\enquote{\textcolor{darkgreen}{Green}}: ours, \enquote{\textcolor{blue1}{Blue}}: others. 
VGG19 (image) signifies a unimodal approach utilizing solely image data. 
\enquote{$||$} and \enquote{$+$} represents \emph{concatenation} and \emph{linear addition}, respectively, used for fusion.
Refer to Tables \ref{tab:comparison}, \ref{tab:model_ablation} of the main paper.
}
\label{fig:hitRatio}
\end{figure*}



\vspace{.2cm}
\noindent
\textbf{Analysis using Hit Ratio (${\cal{H}}it$):} 
We denote ${\cal{H}}it$ evaluation metric as the ratio of the correctly predicted first genre (the dominant genre with the highest confidence score) sample count to the total number of samples in the test set.
Fig. \ref{fig:hitRatio} presents the performance of all multi-modal (image + text) methods using ${\cal{H}}it$. 
Among contemporary deep learning methods, ${\cal{H}}it$ ranged from 58.03\% to 77.41\%;  
while GMU \cite{ICLR_GMU} and CentralNet \cite{CentralNet_ECCV} achieved 78.03\% and 80.54\% ${\cal{H}}it$, respectively. 
Notably, BLIP \cite{BLIP} and CLIP \cite{CLIP} exhibited impressive ${\cal{H}}it$, i.e., 86.12\% and 86.39\%, respectively. 
However, our framework outperformed all others, achieving the highest ${\cal{H}}it$ of 87.55\%, with its variants achieving ${\cal{H}}it$ ranging from 85.17\% to 87.55\%. 
Furthermore, we observed that increasing the number of $MSA$ in $SMSAM$ led to improvements in the ${\cal{H}}it$ metric, highlighting the efficacy of our approach in enhancing performance.

\section{Conclusion}
\label{5sec:conclusion}
\noindent
In this paper, we focused on leveraging visual and textual information from movie posters for multi-label movie genre classification. We did not use other movie-related data such as trailers, clips, plot summaries, etc. We initially introduced $MCAM$ (Multi-head Cross Attention Module), subsequently followed by $SMSAM$ (Sequential Multi-head Self Attention Module), to learn the relations among genres and employed asymmetric loss to handle multi-label classification. We utilized a dataset comprised of 13882 movie posters sourced from IMDb. Our models demonstrated promising performances, outperforming major state-of-the-art architectures. Our findings indicate that relying solely on textual information from movie posters for genre classification yields subpar results. However, integrating textual information with visual information significantly enhances classification. 
In the future, we will endeavor to focus on improving the performance of specific genres: biography, mystery, and fantasy, where our current approach has exhibited average results.

\section*{Acknowledgment}
The authors acknowledge the interns/ researchers who helped with some data preprocessing.

\balance 

\bibliographystyle{IEEEtran}  
\bibliography{ref.bib} 

\begin{thebibliography}{10}
\providecommand{\url}[1]{#1}
\csname url@samestyle\endcsname
\providecommand{\newblock}{\relax}
\providecommand{\bibinfo}[2]{#2}
\providecommand{\BIBentrySTDinterwordspacing}{\spaceskip=0pt\relax}
\providecommand{\BIBentryALTinterwordstretchfactor}{4}
\providecommand{\BIBentryALTinterwordspacing}{\spaceskip=\fontdimen2\font plus
\BIBentryALTinterwordstretchfactor\fontdimen3\font minus \fontdimen4\font\relax}
\providecommand{\BIBforeignlanguage}[2]{{%
\expandafter\ifx\csname l@#1\endcsname\relax
\typeout{** WARNING: IEEEtran.bst: No hyphenation pattern has been}%
\typeout{** loaded for the language `#1'. Using the pattern for}%
\typeout{** the default language instead.}%
\else
\language=\csname l@#1\endcsname
\fi
#2}}
\providecommand{\BIBdecl}{\relax}
\BIBdecl

\bibitem{intro_1}
P.~Winoto \emph{et~al.}, ``The role of user mood in movie recommendations,'' \emph{Expert Systems with Applications}, vol.~37, no.~8, pp. 6086--6092, 2010.

\bibitem{intro_2}
M.~Frey, \emph{Netflix recommends: algorithms, film choice, and the history of taste}.\hskip 1em plus 0.5em minus 0.4em\relax Univ of California Press, 2021.

\bibitem{trailer_1}
R.~M.-Lezama \emph{et~al.}, ``Improving transfer learning for movie trailer genre classification using a dual image and video transformer,'' \emph{Information Processing \& Management}, vol.~60, no.~3, p. 103343, 2023.

\bibitem{trailer_2}
A.~Yadav and D.~K. Vishwakarma, ``A unified framework of deep networks for genre classification using movie trailer,'' \emph{Applied Soft Computing}, vol.~96, p. 106624, 2020.

\bibitem{trailer_3}
G.~S. Simoes \emph{et~al.}, ``Movie genre classification with convolutional neural networks,'' in \emph{IJCNN}, 2016, pp. 259--266.

\bibitem{plot_1}
Q.~Hoang, ``Predicting movie genres based on plot summaries,'' \emph{arXiv:1801.04813}, 2018.

\bibitem{plot_2}
J.~Wehrmann \emph{et~al.}, ``Self-attention for synopsis-based multi-label movie genre classification,'' in \emph{FLAIRS Conf.}, 2018.

\bibitem{plot_3}
S.~Kar \emph{et~al.}, ``{F}olksonomication: Predicting tags for movies from plot synopses using emotion flow encoded neural network,'' in \emph{COLING}, 2018, pp. 2879--2891.

\bibitem{poster_1}
J.~A. Wi \emph{et~al.}, ``Poster-based multiple movie genre classification using inter-channel features,'' \emph{IEEE Access}, vol.~8, pp. 66\,615--66\,624, 2020.

\bibitem{poster_2}
W.-T. Chu and H.-J. Guo, ``Movie genre classification based on poster images with deep neural networks,'' in \emph{MUSA2}, 2017, pp. 39--45.

\bibitem{poster_3}
M.~Pobar \emph{et~al.}, ``Multi-label poster classification into genres using different problem transformation methods,'' in \emph{CAIP}, 2017, pp. 367--378.

\bibitem{clip_1}
X.~Yuan \emph{et~al.}, ``Automatic video genre categorization using hierarchical {SVM},'' in \emph{ICIP}, 2006, pp. 2905--2908.

\bibitem{plot_4}
P.~Gorinski \emph{et~al.}, ``What’s this movie about? a joint neural network architecture for movie content analysis,'' in \emph{NAACL}, 2018, pp. 1770--1781.

\bibitem{multimodel_2}
P.~C.-Bonilla \emph{et~al.}, ``Moviescope: Large-scale analysis of movies using multiple modalities,'' \emph{arXiv:1908.03180}, 2019.

\bibitem{multimodal_1}
R.~B. Mangolin \emph{et~al.}, ``A multimodal approach for multi-label movie genre classification,'' \emph{Multimedia Tools and Applications}, vol.~81, no.~14, pp. 19\,071--19\,096, 2022.

\bibitem{visual_1}
H.~Zhou \emph{et~al.}, ``Movie genre classification via scene categorization,'' in \emph{ACM Multimedia}, 2010, pp. 747--750.

\bibitem{trailer_4}
J.~Wehrmann and R.~C. Barros, ``Movie genre classification: A multi-label approach based on convolutions through time,'' \emph{Applied Soft Computing}, vol.~61, pp. 973--982, 2017.

\bibitem{plot_5}
A.~M. Ertugrul and P.~Karagoz, ``{Movie genre classification from plot summaries using bidirectional LSTM},'' in \emph{ICSC}, 2018, pp. 248--251.

\bibitem{plot_6}
V.~Battu \emph{et~al.}, ``Predicting the genre and rating of a movie based on its synopsis,'' in \emph{PACLIC}, 2018.

\bibitem{ICLR_GMU}
J.~Arevalo \emph{et~al.}, ``Gated multimodal units for information fusion,'' in \emph{ICLR Workshops}, 2017.

\bibitem{CentralNet_ECCV}
V.~Vielzeuf \emph{et~al.}, ``{CentralNet: a Multilayer Approach for Multimodal Fusion},'' in \emph{ECCV Workshops}, 2018.

\bibitem{multimodal_3}
I.~R. Bribiesca \emph{et~al.}, ``Multimodal weighted fusion of transformers for movie genre classification,'' in \emph{MAI-Workshop}, 2021, pp. 1--5.

\bibitem{multilabel_classification}
F.~Herrera \emph{et~al.}, \emph{Multilabel classification}.\hskip 1em plus 0.5em minus 0.4em\relax Springer, 2016.

\bibitem{gemini}
G.~Team \emph{et~al.}, ``Gemini: A family of highly capable multimodal models,'' \emph{arXiv:2312.11805}, 2023.

\bibitem{CLIP}
A.~Radford \emph{et~al.}, ``Learning transferable visual models from natural language supervision,'' in \emph{ICML}, 2021, pp. 8748--8763.

\bibitem{alignment}
C.~Zhang \emph{et~al.}, ``Multimodal intelligence: Representation learning, information fusion, and applications,'' \emph{IEEE Journal of Selected Topics in Signal Processing}, vol.~14, no.~3, pp. 478--493, 2020.

\bibitem{Deep_learning}
A.~Zhang \emph{et~al.}, \emph{Dive into deep learning}.\hskip 1em plus 0.5em minus 0.4em\relax Cambridge Uni. Press, 2023.

\bibitem{attention}
A.~Vaswani \emph{et~al.}, ``Attention is all you need,'' \emph{NeurIPS}, vol.~30, 2017.

\bibitem{asl_cvpr}
T.~Ridnik \emph{et~al.}, ``Asymmetric loss for multi-label classification,'' in \emph{CVPR}, 2021, pp. 82--91.

\bibitem{eval_metrics_acm}
E.~Gibaja and S.~Ventura, ``A tutorial on multilabel learning,'' \emph{ACM CSUR}, vol.~47, no.~3, pp. 1--38, 2015.

\bibitem{BLIP}
J.~Li \emph{et~al.}, ``{BLIP: Bootstrapping language-image pre-training for unified vision-language understanding and generation},'' in \emph{ICML}, 2022, pp. 12\,888--12\,900.

\bibitem{VGG19_ICLR}
K.~Simonyan and A.~Zisserman, ``Very deep convolutional networks for large-scale image recognition,'' \emph{ICLR}, 2015.

\bibitem{ResNet50V2_ECCV}
K.~He \emph{et~al.}, ``Identity mappings in deep residual networks,'' in \emph{ECCV}, 2016, pp. 630--645.

\bibitem{DenseNet121_CVPR}
G.~Huang \emph{et~al.}, ``Densely connected convolutional networks,'' in \emph{CVPR}, 2017, pp. 4700--4708.

\bibitem{MobileNetV2_CVPR}
M.~Sandler \emph{et~al.}, ``Mobilenetv2: Inverted residuals and linear bottlenecks,'' in \emph{CVPR}, 2018, pp. 4510--4520.

\bibitem{InceptionV3_CVPR}
C.~Szegedy \emph{et~al.}, ``Rethinking the inception architecture for computer vision,'' in \emph{CVPR}, 2016, pp. 2818--2826.

\bibitem{EfficientNetV2B2_ICML}
M.~Tan and Q.~Le, ``{EfficientNetV2: Smaller Models and Faster Training},'' in \emph{ICML}, 2021, pp. 10\,096--10\,106.

\bibitem{Word2Vec_1}
T.~Mikolov \emph{et~al.}, ``Efficient estimation of word representations in vector space,'' \emph{ICLR}, 2013.

\bibitem{Word2Vec_2}
T.~Mikolov, I.~Sutskever \emph{et~al.}, ``Distributed representations of words and phrases and their compositionality,'' \emph{NeurIPS}, vol.~26, 2013.

\bibitem{LSTM}
S.~Hochreiter and J.~Schmidhuber, ``Long short-term memory,'' \emph{Neural computation}, vol.~9, no.~8, pp. 1735--1780, 1997.

\end{thebibliography}


\end{document}